\documentclass[aps, onecolumn, preprint, amsmath,amssymb, superscriptaddress, nofootinbib]{revtex4}
\usepackage[pdftex]{graphicx}
 \usepackage{amsmath}
 
\usepackage[T1]{fontenc}
\usepackage{amssymb}
\usepackage{amsfonts}
\usepackage{bm}
 \usepackage{amsmath} 
  \usepackage{flushend}
\usepackage{color}
\usepackage{lipsum}
\usepackage{amsfonts} 
\usepackage{amssymb, mathrsfs}
\usepackage{braket}
\usepackage{graphicx} 
\usepackage{subfigure}
\usepackage{bbm}
\def\beq{\begin{equation}}
\def\eeq{\end{equation}}
\def\bsp{\begin{split}}
\def\esp{\end{split}}
\def\bea{\begin{eqnarray}}
\def\eea{\end{eqnarray}}
\def\ba{\begin{array}}
\def\ea{\end{array}}

\def\dg{\dagger}

\def\lb{\left(}
\def\rb{\right)}

\def\l.{\left.}
\def\r.{\right.}

\def\ra{\rangle}
\def\la{\langle}

\def\bo{{{\bf k}}}

\begin{document}

\date{\today}
\title{ Laser-Irradiated CrI$_3$: When Chiral Photons Meet Topological Magnons}
\author{S. A. Owerre}
\affiliation{Perimeter Institute for Theoretical Physics, 31 Caroline St. N., Waterloo, Ontario N2L 2Y5, Canada.}

\begin{abstract}
Insulating honeycomb ferromagnet  CrI$_3$ has recently attracted considerable attention  due to its potential use for dissipationless spintronics applications. Recently, topological spin excitations have been observed experimentally in bulk CrI$_3$ by  L.   Chen, et al.  \big[Phys. Rev. X {\bf 8}, 041028 (2018)\big] using  inelastic neutron scattering. This suggest that bulk CrI$_3$  has strong spin-orbit coupling and its spin Hamiltonian  should include a next-nearest neighbour Dzyaloshinskii-Moriya  (DM) interaction.    Inspired by this experiment, we study non-equilibrium emergent photon-dressed topological spin and thermal Hall transports in laser-irradiated CrI$_3$ with and without the DM interaction. We show that the spin excitations  can be manipulated into different topological phases with different  Chern numbers.  Most importantly, we show that the emergent photon-dressed spin and thermal Hall response can be switched to different signs. Hence, the generated magnon spin photocurrents can be manipulated  by the laser field, which is of great interest in ultrafast spin current generation and could pave the  way for  future applications of CrI$_3$ to  topological opto-spintronics and opto-magnonics.
\end{abstract}
\vspace{10px}

\maketitle

\section{Introduction}
In traditional modern electronic devices,  Joule heating caused by the flow of electron charge (electric current) is inevitable. It can lead to large information losses.  This  debilitating  problem can be circumvented by utilizing the electron spin degree of freedom which interacts with each other in magnetically ordered materials. In this respect, magnons (quantized collective excitations of electron spins)  play a pivotal role as the primary underlying magnetic spin excitations in insulating magnetically ordered materials. They are neutral quasiparticles and carry an intrinsic spin of $1$ and a spin magnetic dipole moment $\sim g\mu_B\sigma$, where $g$ is the spin g-factor and  $\mu_B$ is the Bohr magneton with $\sigma = \pm 1$. Magnon spintronics  aims  to eliminate the debilitating  problem posed by modern electronic devices by utilizing the spin current carried  by magnons as information carriers \cite{magn1, magn1a, magn2, magn3, magn4, magn5, magn6, magn7, magn1b, magn1c}.

The recent discovery of intrinsic ferromagnetism in van der Waals crystals CrI$_3$ down to monolayer limit \cite{d0,d1} has garnered considerable interest in potential two-dimensional (2D) magnet-based applications, such as ultrathin magnetic sensors and high-efficiency spin filter devices \cite{d2,d3,d4,d5,d6,d7,d8}. Recently, the magnetic spin excitations in the bulk ($T_c \sim 61~{\rm K}$) and monolayer ($T_c \sim 45~{\rm K}$)   CrI$_3$ have been measured by inelastic neutron scattering \cite{cr} and Raman spectroscopy \cite{cr1} respectively at zero magnetic field. Both experiments have identified two distinct spin wave (magnon) modes which constitute the underlying magnetic excitations in   CrI$_3$. In the bulk structure, the two spin wave  modes are separated by a finite energy gap $\Delta_{\bf K}\sim 4~{\rm meV}$ at the Dirac points \cite{cr}, while the magnon bandwidth is $\Delta_{\bf \Gamma}\sim 19~{\rm meV}$. The gap at the Dirac points can only be explained  by the presence of the Dzyaloshinskii-Moriya  interaction (DM) interaction \cite{owerre, dm,dm1}, which breaks the inversion symmetry of the lattice resulting in the time-reversal symmetry breaking of the spin wave modes. The consequence of this result is that the spin wave modes now carry finite Berry curvatures and Chern number protected  edge modes. Therefore, thermal Hall effect \cite{th1,th2,th3,th4,th5,th6,th7} can be manifested in bulk   CrI$_3$. 

The manipulation of magnon spin currents \cite{cra,ruc,pro} is essentially difficult at equilibrium, which limits the potential practical applications of topological insulating magnets in spintronics. Recently, laser-irradiation  of solid-state materials  has attracted considerable attention  as an alternative way for engineering topological nontrivial states from  topologically trivial  quantum materials \cite{pho1, pho2, pho3, pho4, pho5, pho6, pho7, pho7a, pho7b, pho7c}. However, there are very limited applications to real materials \cite{pho8, pho9, pho10}. This formalism, however, depends on coupling between the charge degree of freedom and the electric field component of the laser light through the Peierls phase, which result in non-equilibrium  Floquet-engineered topological phases. In insulating magnets, the spin degree of freedom can couple to the electric field component of the laser light through different processes \cite{spin1,spin2,spin3,spin4,spin5, spin6, spin1a, spin5a,sowe, kar, ely} such as the electric polarization \cite{spin3} or the  Aharonov-Casher phase \cite{aha}, where the former is encoded in the latter.  In this case, the resulting Floquet physics can renormalize the underlying spin Hamiltonian to stabilize magnetic phases and induce  Floquet topological spin excitations.  This  provides a promising avenue for  ultrafast spin current generation. Therefore, laser-irradiated  CrI$_3$ could play an essential role  in ultrafast magnon spin current switching, which could provide a great avenue to opto-spintronics and opto-magnonics \cite{magn1, magn1a, magn2, magn3, magn4, magn5, magn6, magn7, magn1b, magn1c}.  
 
 In this paper,  we study emergent photon-dressed topological spin and thermal Hall transport properties in  laser-irradiated CrI$_3$. We consider the three-dimensional (3D) Heisenberg spin Hamiltonian for bulk CrI$_3$  motivated by recent inelastic neutron scattering  experiment \cite{cr}, with up to third nearest-neighbours  $J_1 = 2.01$ meV, $J_2 = 0.16$ meV, and $J_3 = 0.18$ meV; an easy-axis anisotropy $K = 0.22$ meV, a DM interaction $D = 0.31$ meV, and an interlayer-exchange interaction  $J_c = 0.59$ meV.  We show that in the presence of a laser electric field with varying intensity $E_0$ of order $10^{6}~{\rm V/m}$ and off-resonant photon energy of $\hbar\omega = 30~{\rm meV}$, the spin excitations in  CrI$_3$ can be manipulated into different photon-dressed topological phases with different  Chern numbers. We further show that  the spin and thermal Hall transports can be switched by the polarized photons. The direct implication of this switching is that the magnon spin photocurrents can be manipulated by polarized photons, which could lead to ultrafast spin-switching in  CrI$_3$. We believe that the current results  will provide an essential guide towards ultrafast spin transport properties in laser-irradiated CrI$_3$.

 \begin{figure}
\centering
\includegraphics[width=.8\linewidth]{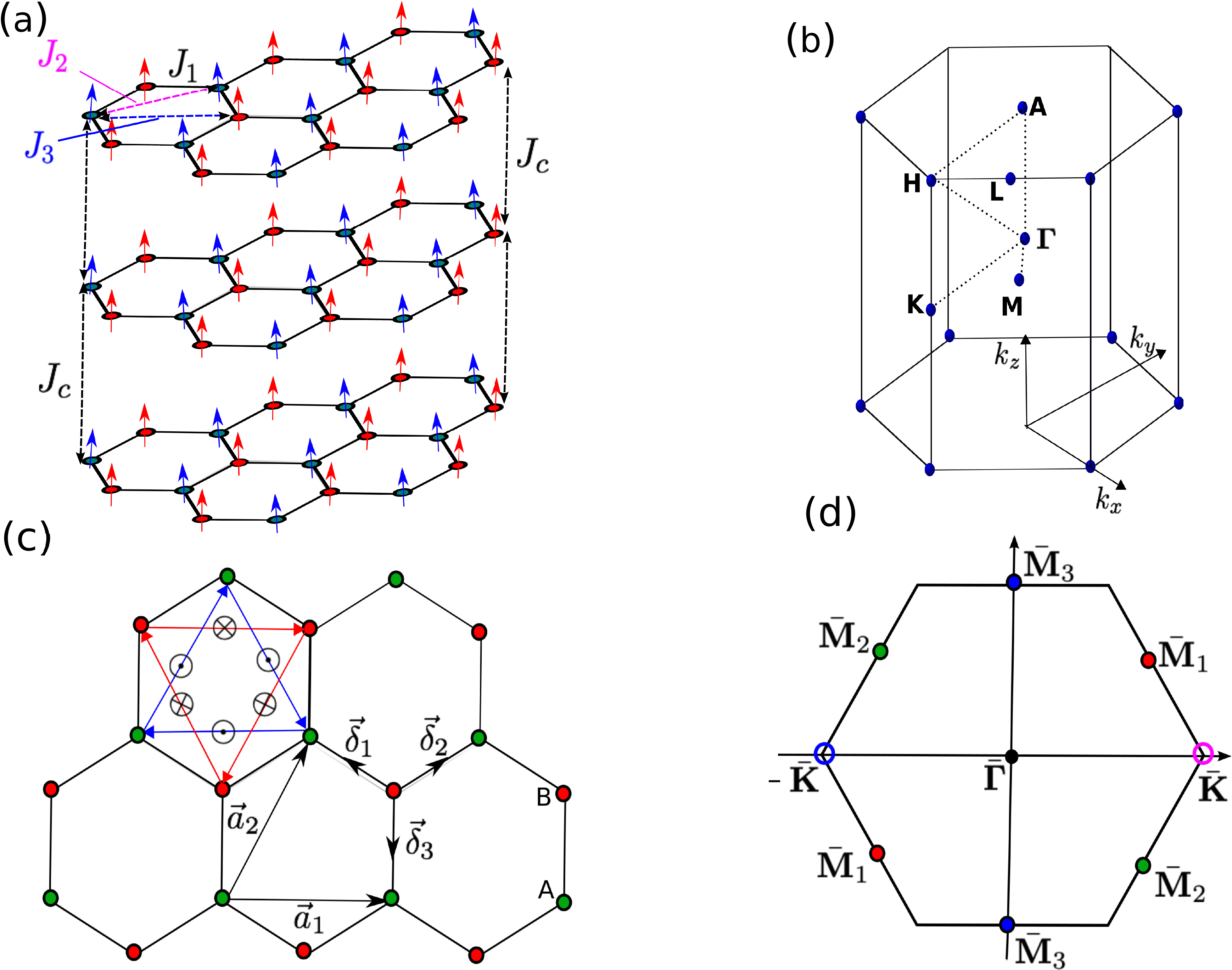}
\caption{Color online. (a) Bulk magnetic structure of  CrI$_3$ with honeycomb-lattice  ferromagnets stacked along the $z$ direction.  (b,d) Bulk Brillouin zone (BZ) of the lattice with high symmetry points. (c) Top view of the bulk honeycomb-lattice structure. The blue and red triangular arrows indicate the DM vector at the midpoint of the bonds. (d) Projected   (001) surface BZ of the hexagonal lattice with high symmetry points.}
\label{lattice}
\end{figure}

\section{Heisenberg spin Hamiltonian}
Based on recent inelastic neutron scattering experiment \cite{cr}, the equilibrium Heisenberg spin Hamiltonian that describes the bulk honeycomb ferromagnet CrI$_3$  can be written as
\begin{align}
\mathcal H&=-\sum_{ \la ij\ra,\ell}J_{ij} {\bf S}_{i}^{\ell}\cdot{\bf S}_{j}^{\ell}+\sum_{ \la\la ij\ra\ra,\ell}{\bf D}_{ij}\cdot{\bf S}_{i}^{\ell}\times{\bf S}_{j}^{\ell}-K\sum_{i,\ell}\big(S_{i}^{z,\ell}\big)^2-J_c\sum_{\la \ell \ell^\prime\ra, i} {\bf S}_{i}^{\ell}\cdot{\bf S}_{i}^{\ell^\prime},
\label{crmodel}
\end{align}
where ${\bf S}_{i}^{\ell}$ is the spin vector at site $i$ in layer $\ell$. The first term is the intralyer Heisenberg ferromagnetic exchange interaction up to third nearest-neighbours $J_{ij}=J_1,J_2,J_3$ as depicted in Fig.~\ref{lattice}(a). The second term is the DM interaction due to lack of inversion symmetry at the midpoint of the second-nearest-neighbour bonds as depicted in Fig.~\ref{lattice}(c). Here $ {\bf D}_{ij}=\nu_{ij}D{\bf \hat z}$, where $\nu_{ij}=\pm 1$ for clockwise and counter-clockwise hopping spin magnetic moments  on each honeycomb layer sublattice.  The third term is the  easy-axis anisotropy, which forces the spins to align along the $z$ direction in the absence of an external magnetic field.   This is in stark contrast to the in-plane  kagome ferromagnet Cu(1,3-bdc) \cite{th2, rchi}, where an external magnetic field is required to polarize the spins along the $z$ axis.  The fourth term is the nearest-neighbour  interlayer ferromagnetic  interaction between the layers.  The experimentally deduced  parameters are \cite{cr}: $J_1 = 2.01$ meV, $J_2 = 0.16$ meV, $J_3 = 0.18$ meV, $D = 0.31$ meV, $K = 0.22$ meV, and $J_c = 0.59$ meV.

 \begin{figure}
\centering
\includegraphics[width=1\linewidth]{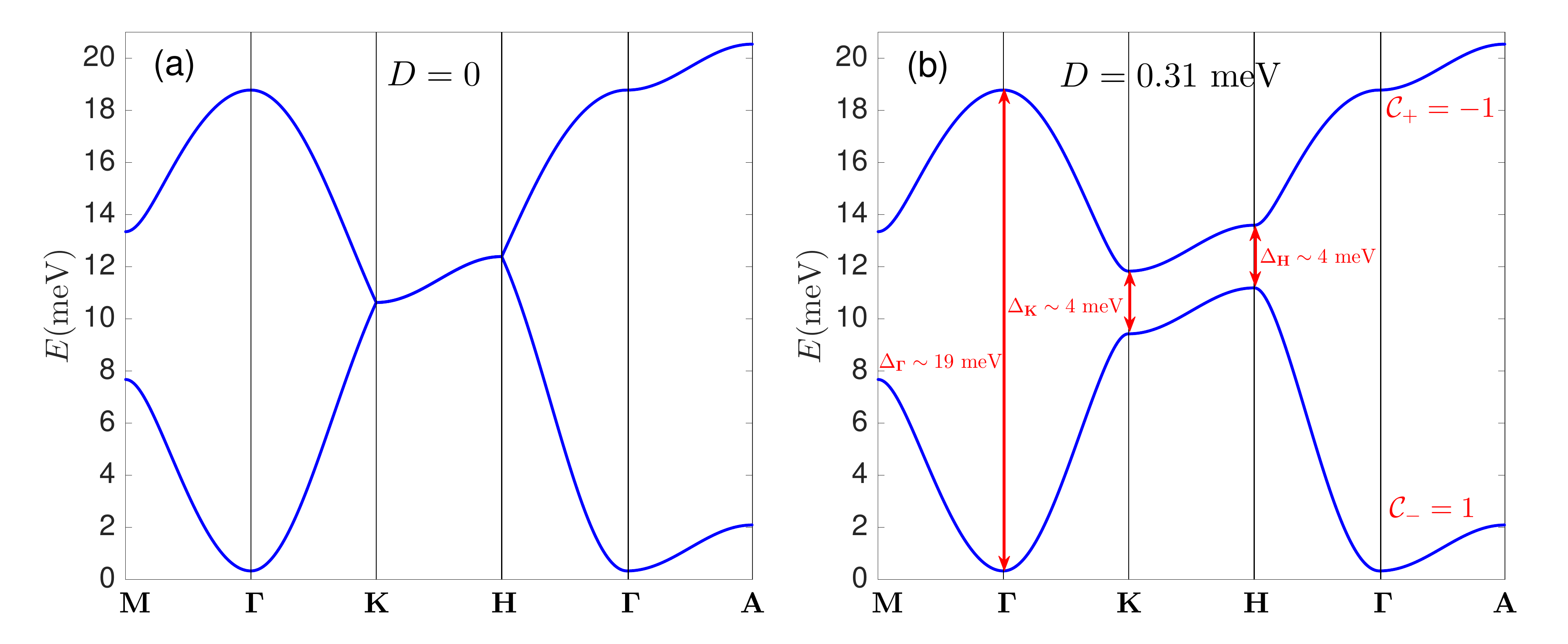}
\caption{Color online.    Equilibrium topological magnon band structure of bulk CrI$_3$ for $D=0$ (a) and $D=0.31~{\rm meV}$ (b). The gap at the Dirac points is $\Delta_{\bf {K}}=\Delta_{\bf {H}}\sim 4~{\rm meV}$ in accordance with the experimentally observed Dirac gap. The magnon bandwidth is $\Delta_{\bf {\Gamma}}\sim 19~{\rm meV}$}.
\label{band}
\end{figure}

\section{Theory of laser-irradiated Honeycomb Ferromagnet CrI$_3$}
In the presence of an intense laser light,  electron spin degree of freedom in magnetic insulators can couple to  laser field  through different processes \cite{spin1,spin2,spin3,spin4,spin5, spin6, spin1a, spin5a, sowe,kar, ely}. In particular, quantized spin waves (magnons) are the oscillations of the spin magnetic dipole moment of an electron. We choose the quantization of the spin magnetic dipole moment along the $z$ direction such that ${\boldsymbol \mu}_S= -g\mu_B\sigma {\bf \hat z}$ with $\sigma = \pm 1$. In the presence of a laser (electric) field ${\bf E}(\tau)$, the hopping spin magnetic dipole moments accumulate a time-dependent version of the Aharonov-Casher effect \cite{aha}
\begin{align}
\Phi_{ij}(\tau) = \frac{g\mu_B}{\hbar c^2}\int_{{\bf r}_i}^{{\bf r}_j}{\boldsymbol \Xi}(\tau) \cdot d{\boldsymbol \ell},
\label{magn}
\end{align}
where $\sigma = 1$ has been used for ferromagnetic spins. ${\boldsymbol \Xi}(\tau)= {\bf E}(\tau) \times {\bf \hat z}$ with ${\bf E}(\tau)=-\partial_\tau \bf{A}(\tau)$, where $\bf{A}(\tau)$ is the time-dependent vector potential of the applied laser field. Here $c$ is the speed of light and $\hbar=h/2\pi$ is the reduced Plank's constant. We note that the electric field of the laser light ${\bf E}(\tau)$  can also couple to spin degree of freedom through the electric polarization (magnetoelectric coupling) as in multiferroics \cite{spin1,spin2,spin3}.  In both cases, the coupling between the spin degree of freedom and the electric field of the laser light can result in a laser-induced scalar spin chirality (or spin-orbit interaction) when only first nearest-neighbour hopping is considered.  

As an alternative method, we can consider the  charge degree of freedom, which also couples to a laser field as in Mott insulators \cite{step,claas}. In this case, the electrons accumulate the Peierls phase 
\begin{align}
\varphi_{ij}(\tau) =\frac{e}{\hbar }\int_{{\bf r}_i}^{{\bf r}_j}{\bf A}(\tau) \cdot d{\boldsymbol \ell},
\label{mott}
\end{align}
where $e$ is the electron charge. We note that both phases in Eqs.~\eqref{magn} and \eqref{mott} originate from the same space-homogeneous electric field of a laser light. The dimensionless quantities that characterize the light intensity in Eqs.~\eqref{magn} and \eqref{mott}  are  $\alpha_m = g\mu_BE_0a/\hbar c^2$ and  $\alpha_e = eE_0a/\hbar \omega$ respectively, where $a$ is the lattice constant. Therefore,  the duality  relation $\alpha_m=\alpha_e$ yields

 \bea 
 g\mu_B = \frac{ec\lambda}{2\pi},
 \label{dual}
 \eea
where $\lambda$ is the light wavelength. We note that  Eq.~\eqref{dual} is only valid for time-dependent electric field of a laser light. It does not hold for spatially varying time-independent electric field \cite{hagen}.  
 
 The relation in Eq.~\eqref{dual} implies that the spin magnetic dipole moment $g\mu_B$ carried by magnons in periodically driven magnetic insulators can be tuned by the light wavelength $\lambda$. In fact, it is comparable to the electron charge $e$ for a typical experimentally feasible light wavelength  of order $10^{-8}$m. This shows that the two formalisms can be equivalent. In fact, the resulting effect of  both phases in Eqs.~\eqref{magn} and \eqref{mott} on magnetic insulators  with only first nearest-neighbour hopping parameter is a laser-induced scalar spin chirality or DM interaction \cite{step,claas,sowe}.  

\subsection{Light propagation along the ${\bf z}$-direction}
For light propagation along the $z$-direction, we choose the time-periodic vector potential $\bf{A}(\tau)$  such that the corresponding time-dependent electric field is given by
\begin{align}
& {\boldsymbol \Xi}(\tau)=E_0\big[\sin(\omega \tau),\sin(\omega \tau + \phi), 0\big],
\label{osci}
 \end{align}
 where $E_0$ is the amplitude of the electric field, $\omega$ is the angular frequency of light and $\phi$ is the phase difference. For circularly-polarized light $\phi =\pi/2$ and for linearly-polarized light $\phi=0$ or $\pi$.
 
The electron spin magnetic dipole couples to the laser electric field through the Aharonov-Casher phase, in the same way the electron charge couples through the Peierls phase. Thus, the resulting time-dependent Heisenberg spin Hamiltonian is given by
\begin{align}
\mathcal H(\tau)&=-\sum_{ ij,\ell}J_{ij} \Big[ S_{i}^{z,\ell}S_{j}^{z,\ell}+\frac{1}{2}\big(S_{i}^{+,\ell}S_{j}^{-,\ell}e^{i\Phi_{ij,\ell}(\tau)} + {\rm H.c.}\big)\Big]\label{eqn6}\nonumber\\&+\frac{D}{2}\sum_{ \la\la ij\ra\ra,\ell} {\nu}_{ij,\ell}\big(iS_{i}^{+,\ell}S_{j}^{-,\ell}e^{i\Phi_{ij,\ell}(\tau)} + {\rm H.c.}\big)\nonumber\\&-K\sum_{i,\ell}\big(S_{i}^{z,\ell}\big)^2-J_c\sum_{\la \ell \ell^\prime\ra, i} {\bf S}_{i}^{\ell}\cdot{\bf S}_{i}^{\ell^\prime},
\end{align}
where $S^{\pm,\ell}_{i}= S^{x,\ell}_{i} \pm i S^{y,\ell}_{i}$ denote the spin raising and lowering  operators. Note that the interlayer coupling is not affected by light propagation along the $z$-direction.    The spin current can be derived as $J^S = \partial \mathcal H (\tau)/\partial \Phi_{ij,\ell} (\tau)\equiv\sum_{j\in i}j^s_{ij,\ell}$. For $D=0$, the spin current is given by $j^s_{ij,\ell} = -i\frac{J_{ij}}{2}e^{i\Phi_{ij,\ell} (\tau)}S_i^{-,\ell}S_j^{+,\ell} + {\rm H.c.}$ Thus, the time-dependent Aharonov-Casher phase $\Phi_{ij}(\tau)$ acts as a vector potential or gauge field to the spin current. In the off-resonant limit, when the photon energy is greater than the energy scale of the static system, the effective static Hamiltonian is given by \cite{pho4, pho7} $\mathcal H_{eff}\approx \mathcal H_{0}+ \Delta\mathcal H_{eff}$, where $\Delta\mathcal H_{eff}=\big[\mathcal H_{1}, \mathcal H_{-1}\big]/\hbar\omega$  is the photon emission and absorption term. The first term  $\mathcal H_{0}$ is the original time-independent spin Hamiltonian with renormalized interactions $J^\perp_{ij}\to J_{ij}^\perp\mathcal J_0(\alpha_m)$ for $S^xS^x$ and $S^yS^y$ couplings, $J^z_{ij}\to J_{ij}$ for $S^zS^z$ coupling, and $D\to D\mathcal J_0(\alpha_m)$, where  ${\mathcal J}_n(x)$ is the Bessel function of order $n$. For  dominant $J_1$ interaction, the second term yields a photoinduced DM interaction is of the form \cite{spin3, sowe, kar, ely}  $\Delta \mathcal H_{eff}\propto D_F\sum_{ \langle\langle ijk\rangle\rangle} \nu_{jk}{\bf S}_i\cdot\big({\bf S}_j \times {\bf S}_k\big)$, where  ${\bf S}_i=S_i^z{\hat e}_z$, $D_F\propto \sqrt{3}\sin\phi J_1^2{\mathcal J}_1^2({\alpha_m})/\hbar\omega$.  The photoinduced DM interaction arises from the commutation  relation  $\big[ S_\alpha^+ S_\beta^-, S_\rho^+ S_{\gamma}^-\big]=2\big( \delta_{\beta\rho}S_{\beta}^zS_\alpha^+ S_\gamma^- - \delta_{\alpha\gamma}S_{\alpha}^zS_\rho^+ S_\beta^-\big)$, which means that $J_2$ and $J_3$ will also have a small contribution, but the dominant contribution comes from the $J_1$ term.      Thus, the  resulting static effective spin Hamiltonian in the present model comprises a photoinduced DM interaction (scalar spin chirality) and a renormalized intrinsic DM interaction which competes with each other. The spin-wave excitations of this static effective spin Hamiltonian can be captured by performing linear spin wave theory directly from Eq.~\eqref{eqn6} and taking the off-resonant limit. 


We note that below the Curie temperature $T<T_c=  61~{\rm K}$, the spin wave excitations in  CrI$_3$ can be captured in linear spin wave theory by performing the linearized Holstein Primakoff  transformation
\begin{align}
S_{i}^{z,\ell}= S-a_{i}^{\dagger,\ell} a_{i}^{\ell},~S_{i}^{+,\ell}\approx \sqrt{2S}a_{i}^{\ell}=(S_{i}^{-,\ell})^\dg,
\end{align}
where $a_{i}^{\dagger,\ell} (a_{i}^{\ell})$ are the bosonic creation (annihilation) operators. The resulting time-dependent linear  spin-wave Hamiltonian in momentum space can be written as 
 \begin{align}
 \mathcal H (\tau) = E_{MF} + \sum_{{\bf k}} \psi^\dg(\bf{k})\cdot\mathcal H(\bf{k},\tau)\cdot\psi(\bf{k}),
 \end{align}
where $E_{MF}$ is the mean-field energy, $\psi^\dg({ \bf k})=(a_{{\bf k},A}^\dg,a_{ {\bf k},B}^\dg)$ is the basis vector, and $\bo = (k_x,k_y,k_z)$ is the 3D momentum vector.  The momentum space Hamiltonian is given by  
\begin{widetext}
\begin{align}
 \mathcal H({\bf{k}},\tau)=S
 \begin{pmatrix}
\rho(k_z)+\rho_{D}({\bf k}_\parallel,\tau)+\rho_2({\bf k}_\parallel,\tau)&\rho_1({\bf k}_\parallel,\tau)+ \rho_3({\bf k}_\parallel,\tau)\\  \rho_1^\dg({\bf k}_\parallel,\tau)+ \rho_3^\dg({\bf k}_\parallel,\tau)& \rho(k_z)+\rho_{D}(-{\bf k}_\parallel,\tau)+\rho_2({\bf k}_\parallel,\tau)
 \end{pmatrix}.
 \label{hamp}
\end{align}
\end{widetext}
The prefactor  corresponds to  $S=3/2$ for CrI$_3$, and  ${\bf k}_\parallel = (k_x,k_y)$ is the in-plane momentum wave vector.
\begin{align}
&\rho(k_z) = 3(J_1+J_3) +2K+6J_2+2J_c(1- \cos k_z)
\end{align}

\begin{align}
&\rho_1({\bf k}_\parallel,\tau) = -J_1\big[e^{-i\mu_m{\boldsymbol \Xi}(\tau)\cdot{\boldsymbol{\delta}_3}} + e^{-i{\bf k}_{\parallel}\cdot{\bf a}_2} e^{-i\mu_m{\boldsymbol \Xi}(\tau)\cdot{\boldsymbol{\delta}_2}}\\&\nonumber + e^{-i{\bf k}_{\parallel}\cdot{\bf a}_1} e^{-i\mu_m{\boldsymbol \Xi}(\tau)\cdot{\boldsymbol{\delta}_1}}\big],
\end{align}

\begin{align}
&\rho_2({\bf k}_\parallel,\tau)= - J_2\big[e^{-i({\bf k}_{\parallel}+\mu_m{\boldsymbol \Xi}(\tau))\cdot{\bf a}_1} +e^{-i({\bf k}_{\parallel}+\mu_m{\boldsymbol \Xi}(\tau))\cdot{\bf a}_2}\\&\nonumber+e^{-i({\bf k}_{\parallel}+\mu_m{\boldsymbol \Xi}(\tau))\cdot{\bf a}_3}   +\rm{H.c.}\big],
\end{align}

\begin{align}
& \rho_3({\bf k}_\parallel,\tau) = -J_3\big[e^{-i{\bf k}_{\parallel}\cdot({\bf a}_1+{\bf a}_2)} e^{-i\mu_m{\boldsymbol \Xi}(\tau)\cdot 2{\boldsymbol{\delta}_3}}\\&\nonumber 
+e^{i{\bf k}_{\parallel}\cdot{\bf a}_3}e^{-i\mu_m{\boldsymbol \Xi}(\tau)\cdot ({\bf a}_1+{\boldsymbol{\delta}_2})}+e^{-i{\bf k}_{\parallel}\cdot{\bf a}_3}e^{-i\mu_m{\boldsymbol \Xi}(\tau)\cdot ({\bf a}_2+{\boldsymbol{\delta}_1})} \big],
\end{align}

\begin{align}
&\rho_D({\bf k}_\parallel,\tau)= -iD\big[e^{-i({\bf k}_{\parallel}+\mu_m{\boldsymbol \Xi}(\tau))\cdot{\bf a}_1} +e^{i({\bf k}_{\parallel}+\mu_m{\boldsymbol \Xi}(\tau))\cdot{\bf a}_2}\\&\nonumber+e^{i({\bf k}_{\parallel}+\mu_m{\boldsymbol \Xi}(\tau))\cdot{\bf a}_3}-\rm{H.c.}\big],
\end{align}
where $\mu_m = g\mu_B/\hbar c^2$. The nearest-neighbour vectors are given by ${\boldsymbol{\delta}_1} = -\lb \sqrt{3}/2,~1/2\rb a$,  ${\boldsymbol{\delta}_2} = \lb \sqrt{3}/2,~-1/2\rb a$, ${\boldsymbol{\delta}_3} = (0,~1)a$.  The primitive lattice vectors are given by $ {\bf a}_{1}={\boldsymbol{\delta}_3}-{\boldsymbol{\delta}_1}= \lb\sqrt{3}/2,~3/2\rb a$, $ {\bf a}_2={\boldsymbol{\delta}_3}-{\boldsymbol{\delta}_2}=\lb-\sqrt{3}/2,~3/2\rb a$, and ${\bf a}_3= {\bf a}_1-{\bf a}_2$.

Next,  we  apply the Floquet theory (see Appendix). The magnon modes at equilibrium $E_0=0$ are depicted in Fig.~\eqref{band}  with the experimentally deduced parameters for $D=0$ (a) and $D=0.31~{\rm meV}$ (b). In the former,  the acoustic (lower) and the optical (upper) magnon modes form nodal line along the ${\bf K}$-${\bf H}$ line. In the latter, however,  the acoustic and the optical magnon modes are well-separated due to  nonzero DM interaction. In accordance with experimental observation \cite{cr}, the optical magnon mode extends close to $20$ {\rm meV} in the $k_z=0$ plane and the acoustic magnon mode extends up to $2$ {\rm meV} along the line $k_x=k_y=0$. The small gap at the ${\bf \Gamma}$ point is due to nonzero easy-axis anisotropy $K=0.22$ meV.  The gap at the Dirac points at ${\bf K}$ and ${\bf H}$ is $\sim 4$ {\rm meV} as observed in  experiment and can be attributed to a nonzero DM interaction $D=0.31$ meV. The  acoustic and optical magnon modes carry fixed Chern number $\mathcal C= \pm 1$ respectively.

\subsection{Laser-induced topological transitions in CrI$_3$}
In our numerical calculations, we will utilize  the electron spin magnetic dipole moment $\mu_B = e\hbar/2m_e$, where the g-factor has been set to $2$ and $m_e$ is the electron mass.  Hence, the dimensionless intensity of light in the present formalism is $\alpha_m = eE_0a/m_ec^2$.    To enhance the effect of light on the magnon bands, we consider a laser electric field amplitude of order $E_0=\mathcal O(10^6~{\rm V/m})$, and set the lattice constant $a$ to unity. Unless otherwise stated, we will use the experimentally deduced parameters for CrI$_3$ listed above and fix the photon energy to $\hbar\omega = 30~{\rm meV}$, which is greater than the magnon bandwidth of the undriven system. In Floquet topological systems, it is customary to assume that the quasienergy levels of the Floquet Hamiltonian are close to the equilibrium system, which is realized in the off-resonant limit \cite{pho1, pho2, pho3, pho4, pho5, pho6, pho7, pho7a, pho7b, pho7c}.  Therefore, the properties of equilibrium topological systems can be applied to  Floquet topological systems.

First, we consider the laser-induced Chern number topological phase transitions of the system and focus on the lowest acoustic Floquet magnon branch.  Theoretically, we can consider the 3D ferromagnetic  bulk structure of CrI$_3$  as slices of 2D monolayer CrI$_3$  interpolating between the $k_z=0$ and $k_z=\pi$ planes.  For an arbitrary $k_z$ point the Chern number of the magnon energy branches can be defined as
\begin{align}
\mathcal C_{n}^F = \frac{1}{2\pi}\int_{BZ} d{\bf k}_\parallel\Omega_{n,xy}({\bf k}_\parallel,k_z),
\end{align}
where  $\Omega_{n,\alpha\beta}({\bf k}_\parallel,k_z)$ is the momentum space Floquet Berry curvature for a given magnon quasienergy band $n$ defined as

\begin{align}
\Omega_{n,\alpha\beta}(\bo)=  -2{\rm Im}\sum_{m\neq n}\frac{\braket{u_{n}(\bo)|\hat v_\alpha|u_{m}(\bo)}\braket{u_{m}(\bo)|\hat v_\beta|u_{n}(\bo)}}{\big[ \varepsilon_{n}(\bo)-\varepsilon_{m}(\bo)\big]^2},
\label{chern2}
\end{align}
where $\hat v_\alpha=\partial \mathcal{H}_{eff}(\bo)/\partial k_\alpha$ defines the velocity operators with $\alpha\neq \beta= (x,y,z)$ and $\mathcal{H}_{eff}(\bo)$ is the static effective Hamiltonian and $\varepsilon_{m}(\bo)$ are its Floquet quasienergies (see Appendix).  For an arbitrary $k_z$ point the Chern number is well-defined as far as the denominator in Eq.~\eqref{chern2} is nonzero.   The Chern number of all the 2D slices at an arbitrary $k_z$ point is the same as the planes at two momenta $k_z = k_i$ and $k_z = k_i+\delta k_i$  can be adiabatically connected without closing the gap.  

\begin{figure}
\centering
\includegraphics[width=0.8\linewidth]{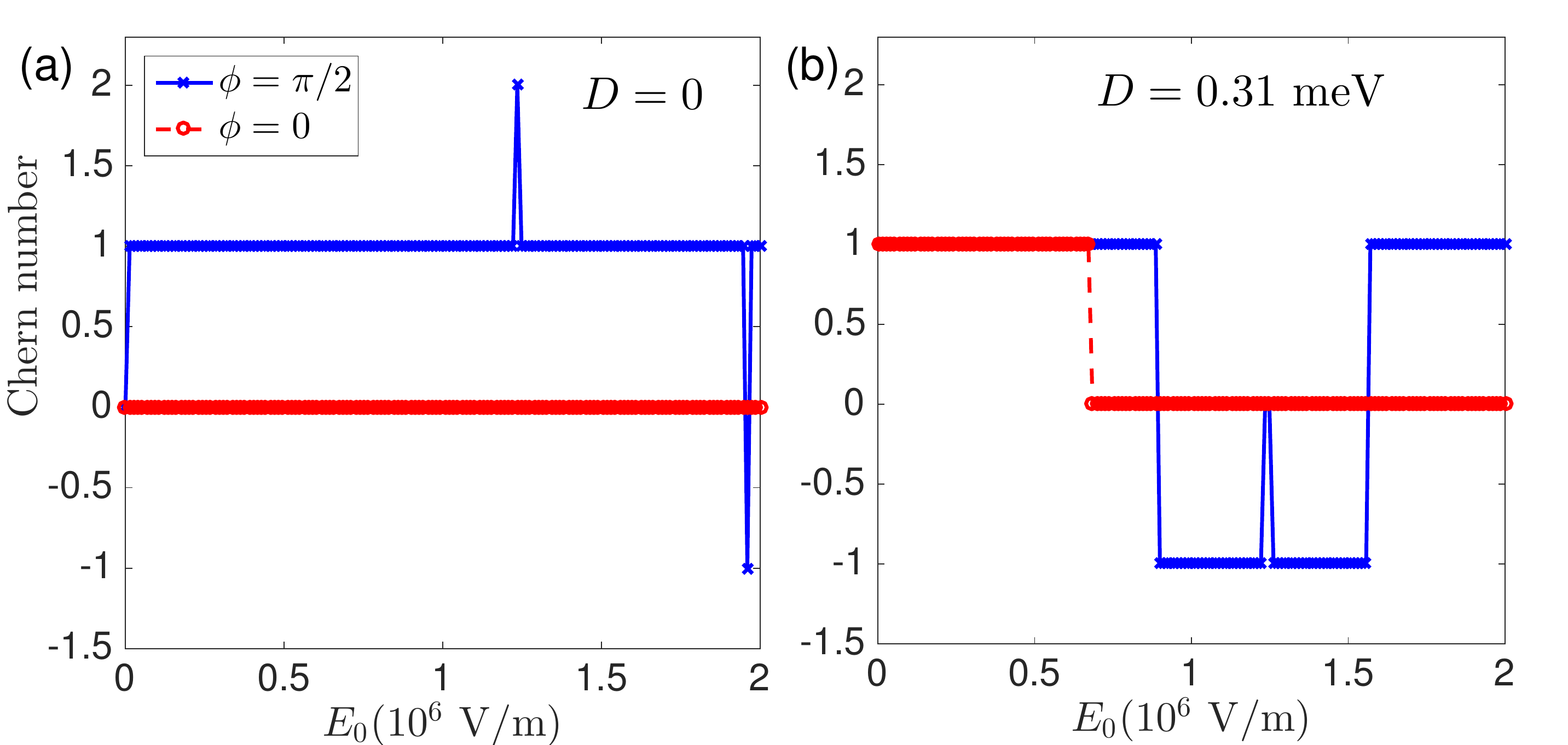}
\caption{Color online.  Laser-induced Chern number topological phase transitions of the lowest  magnon mode in CrI$_3$ as a function of the electric field intensity $E_0$ for $D=0$ (a)  and $D=0.31~{\rm meV}$ (b). The photon energy is $\hbar\omega = 30~{\rm meV}$.}
\label{ChernN}
\end{figure}

\begin{figure}
\centering
\includegraphics[width=0.8\linewidth]{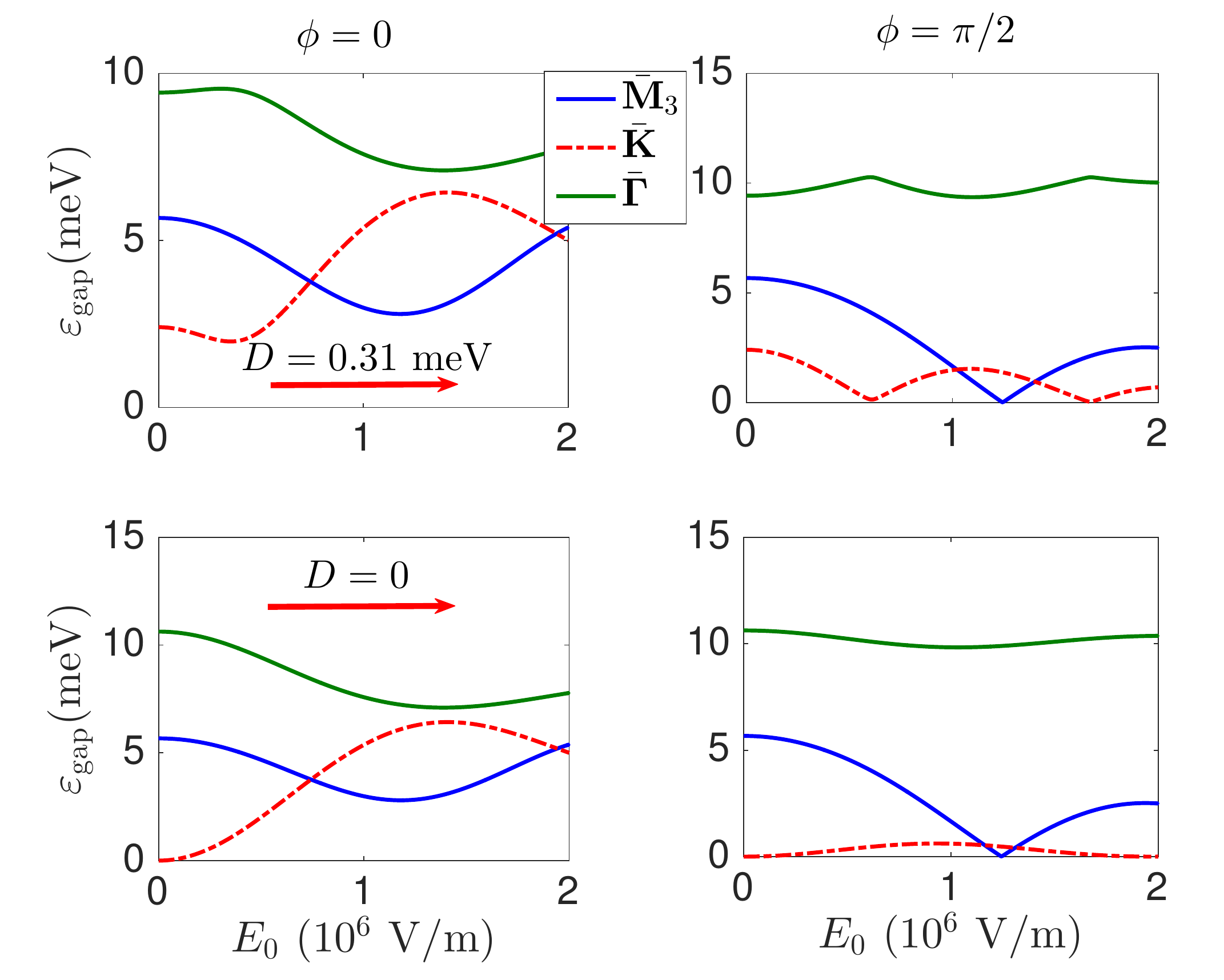}
\caption{Color online. The energy gap between the two Floquet topological magnon modes as a function of electric field intensity $E_0$  on the $k_z=0$ plane for $\phi=0$ left panel and $\phi=\pi/2$ right panel.  The photon energy is $\hbar\omega = 30~{\rm meV}$. }
\label{Gap}
\end{figure}

In Fig.~\eqref{ChernN}, we have shown the evolution of the lowest Floquet Chern number as a function of electric field amplitude $E_0$ for $D=0$ (a) and $D=0.31~{\rm meV}$ (b). In the former, we can see that circularly-polarized light $\phi=\pi/2$ induces  topological transitions in the magnon modes with the Chern number changing  between $-1, +1, +2$ as $E_0$ varies. The sharp jump at $+2$ and $-1$ occur near the zeros of ${\mathcal J}_0$, which renormalizes the spin interactions in the zeroth-order effective spin Hamiltonian $\mathcal H_0$. For linearly-polarized light $\phi=0$ and $D=0$, the system is topologically trivial with zero Chern number because time-reversal symmetry is not broken  for $\phi=0$ and $D=0$.  In the former, however, the system is already topologically nontrivial at equilibrium $E_0=0$ with the Chern number $\mathcal C = +1$ for the lowest magnon mode. But there is no topological phase transition as the parameters are fixed.  As the laser field is turned on,  a topological phase transition occurs  for $\phi = \pi/2$ and $\phi = 0$, with the Chern number changing between $-1,0,+1$. Therefore, the system can be switched from a topologically nontrivial state  to a topologically trivial state and back to a topologically nontrivial state for $\phi =\pi/2$. The distribution of the Chern number is accompanied by a gap closing and reopening at the $\bar{\bf M}_i$ ($i=1,2,3$) and $\bar{\bf K}$ points as we show in Fig.~\eqref{Gap}.

\begin{figure}
\centering
\includegraphics[width=1\linewidth]{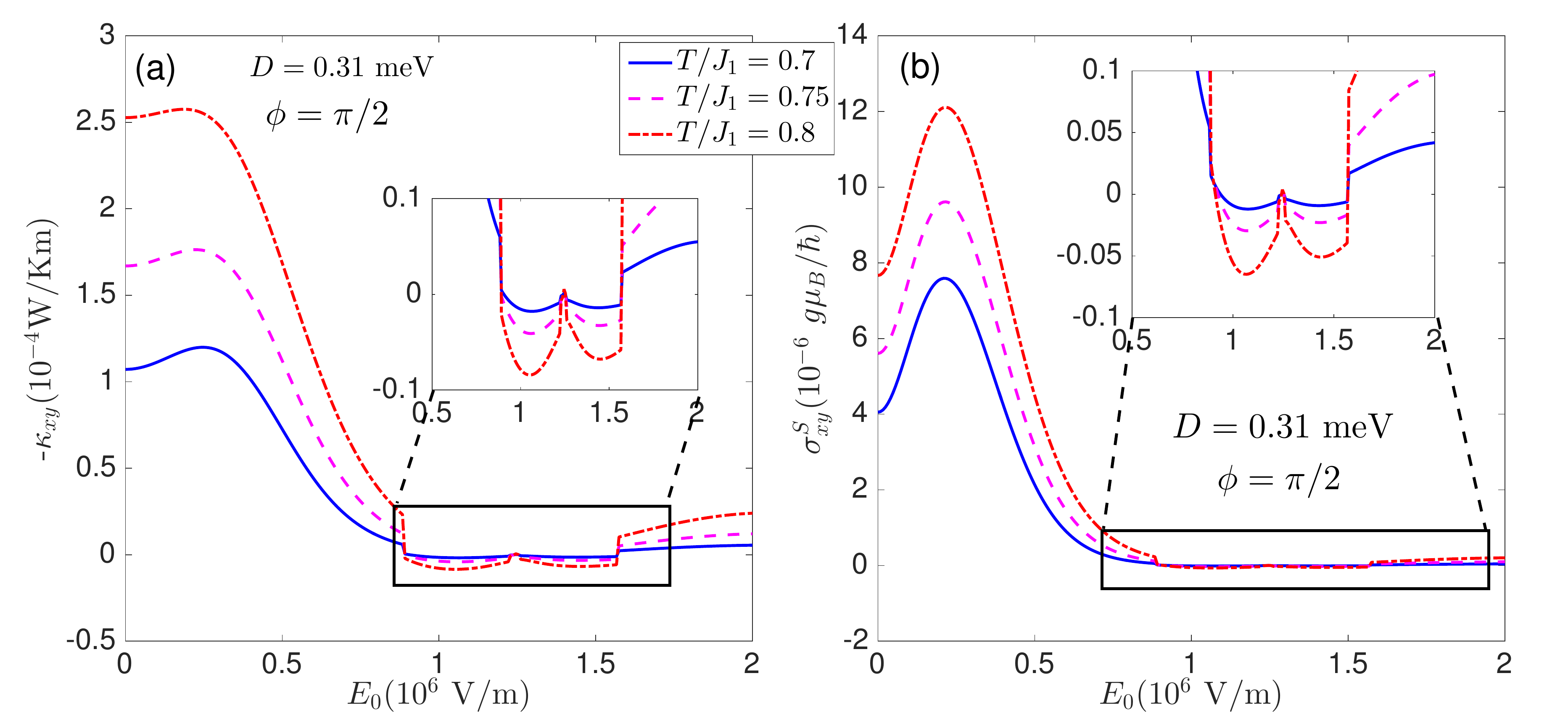}
\caption{Color online. Laser-induced spin and thermal Hall effects in CrI$_3$ as a function of $E_0$  for  $\phi=\pi/2$ and various temperature values. Inset shows the magnified region (cf. Fig.~\eqref{ChernN}(b)).}
\label{THE}
\end{figure}

\begin{figure}
\centering
\includegraphics[width=1\linewidth]{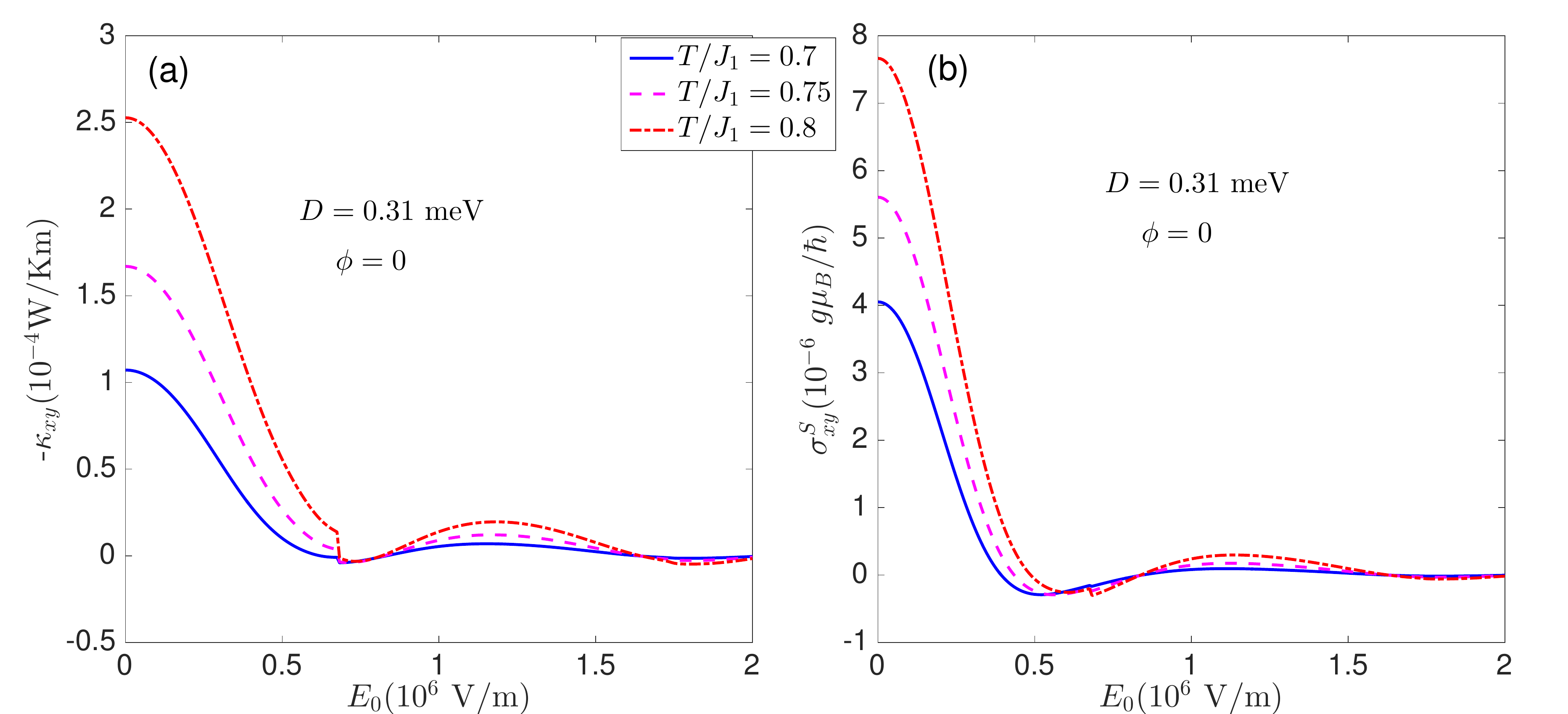}
\caption{Color online. Laser-induced spin and thermal Hall effects in CrI$_3$ as a function of $E_0$  for  $\phi=0$ and various temperature values. }
\label{THE1}
\end{figure}

\begin{figure}
\centering
\includegraphics[width=1\linewidth]{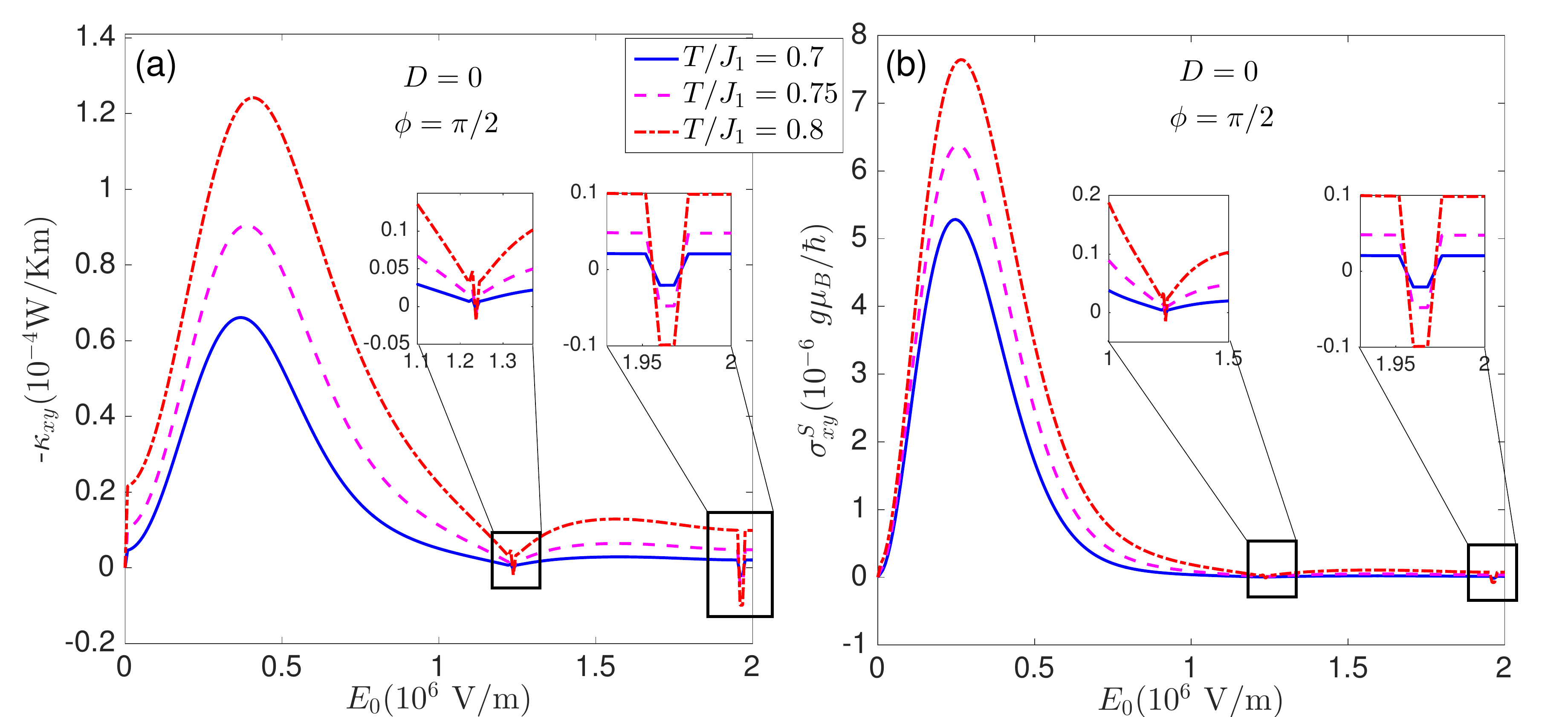}
\caption{Color online.  Laser-induced spin and thermal Hall effects  at zero DM interaction as a function of $E_0$  for  $\phi=\pi/2$ and various temperature values. Inset shows the magnified region (cf. Fig.~\eqref{ChernN}(a)).}
\label{THE0}
\end{figure}

%

\subsection{Laser-induced spin \& thermal Hall effects in CrI$_3$}
The spin and thermal Hall effects are  the hallmarks of topological insulating magnets. They determine the nature of spin excitations in magnetic insulators. For ordered magnets, it is believed that the spin and thermal Hall effects are a consequence of topological spin excitations in equilibrium  systems \cite{th1,th2,th3,th4,th5,th6,th7, she1, she2} . In the non-equilibrium system, we consider the limit when the Bose occupation function is close to thermal equilibrium. Hence,   a longitudinal temperature gradient $-{ \nabla}_\beta { T}$ induces a transverse heat current $ J^Q_{\alpha}=-\sum_{\beta}\kappa_{\alpha\beta}\nabla_{\beta} T$, where $\kappa_{\alpha\beta}$ is the  thermal Hall conductivity \cite{th5} given by

\begin{align}
\kappa_{\alpha\beta} =- k_BT\int_{BZ} \frac{d\bo}{(2\pi)^3}~ \sum_{n=1}^N c_2\big[ f_n^B(\bo)\big]\Omega_{n, \alpha\beta}(\bo),
\label{thm}
\end{align}
where $ f_n^B(\bo)=\big( e^{\varepsilon_{n}(\bo)/k_BT}-1\big)^{-1}$ is the Bose occupation function close to thermal equilibrium, $k_B$ the Boltzmann constant which we will  set to unity, $T$ is the temperature  and $ c_2(x)=(1+x)\lb \ln \frac{1+x}{x}\rb^2-(\ln x)^2-2\text{Li}_2(-x)$ weight function with $\text{Li}_2(x)$ being the  dilogarithm.

A magnetic field gradient ${\nabla}_\beta {B}^z$ can also induce a transverse spin current $ J^S_{\alpha}=\sum_{\beta}\sigma_{\alpha\beta}^S\nabla_{\beta} B^z$, where $\sigma_{\alpha\beta}$ is the spin Hall conductivity \cite{she1,she2} given by

\begin{align}
\sigma_{\alpha\beta}^S =\frac{g\mu_B}{\hbar}\int_{BZ} \frac{d\bo}{(2\pi)^3}~ \sum_{n=1}^N  f_n^B(\bo)\Omega_{n, \alpha\beta}(\bo),
\label{she}
\end{align}

In Fig.~\eqref{THE} and Fig.~\eqref{THE1}, we have shown the trends of the laser-induced spin $\sigma_{xy}^S$ and thermal  $\kappa_{xy}$  Hall conductivities in CrI$_3$ for  $\phi =\pi/2$ and $\phi =0$ respectively.  Note that the dimensionless Curie temperature of CrI$_3$ is $ T_c/J_1 \sim 2.6$, hence the magnon bands are well-defined for $T<T_c$.  Due to nonzero DM interaction, $\kappa_{xy}$ and $\sigma_{xy}^S$  are nonzero at equilibrium $E_0 =0$ with $\kappa_{xy}$ being negative. In the presence of a periodic drive $E_0 \neq 0$, there is a discernible change in both  $\kappa_{xy}$ and $\sigma_{xy}^S$ as $E_0$ varies. In particular,  for  $\phi=\pi/2$, $\kappa_{xy}$ and $\sigma_{xy}^S$ first increase and reach a maximum peak before decreasing rapidly with increasing $E_0$. At low temperature, there is a  sign change in $\kappa_{xy}$ and $\sigma_{xy}^S$  for $\phi=\pi/2$, which  coincides with the change of sign in the Chern number as shown in Fig.~\eqref{ChernN}(b). This shows that the magnon spin photocurrents can be switched by circularly-polarized laser field, which could pave the way towards opto-spintronics and opto-magnonics \cite{magn1, magn1a, magn2, magn3, magn4, magn5, magn6, magn7, magn1b, magn1c}. For  $\phi =0$,  both  $\kappa_{xy}$ and $\sigma_{xy}^S$ decrease rapidly with increasing $E_0$. They eventually hit a bump where the Chern number (Berry curvature) changes to zero. However,  they do not vanish completely in the regime of zero Chern number because of the bosonic nature of magnons. In other words, the Hall response is not a consequence of the Chern number in bosonic systems. It solely depends on the Berry curvature, which can be nonzero even when the Chern number is zero.  In all cases, the Hall response becomes smaller after a topological phase transition. This is due to the change in the Berry curvature when the gap closes and reopens.

For zero DM interaction, $\kappa_{xy}$ and $\sigma_{xy}^S$ vanish at equilibrium $E_0=0$ and for $\phi=0$ because time-reversal symmetry is not broken macroscopically.  For $\phi =\pi/2$, however, time-reversal symmetry is broken and a nonzero $\kappa_{xy}$ and $\sigma_{xy}^S$ is manifested  for $E_0 \neq 0$ as shown in Fig.~\eqref{THE0}. We can see that the sharp drop in $\kappa_{xy}$ and $\sigma_{xy}^S$ is consistent with the jump in the Chern number plot in Fig.~\eqref{ChernN}(a).  

\section{Conclusion}
 
 In summary, we have showed that the spin and thermal Hall response in CrI$_3$ can be switched by a laser field. We also showed that photon-dressed topological phase transitions can occur in the spin wave excitations.  The current results apply to both bulk and monolayer CrI$_3$ with and without the DM interaction.  A direct implication of the current results is that ultrafast spin photocurrents can be generated  in CrI$_3$, which can be studied by the inverse Faraday effect. In addition, terahertz spectroscopy also provides a means to access the photon-dressed topological spin wave band structure in insulating magnets.  Therefore, we have provided a platform for investigation of  new features in CrI$_3$ and its potential applications to opto-spintronics, opto-magnonics, and magnon spintronics \cite{magn1, magn1a, magn2, magn3, magn4, magn5, magn6, magn7, magn1b, magn1c}.

\appendix

\section{Floquet-Bloch theory}
 We implement the Floquet theory to study the periodically driven magnetic insulator CrI$_3$.  Since the  Hamiltonian is time-periodic we can expand it as
\begin{align}
\mathcal H({\bf k},\tau)=\mathcal H({\bf k},\tau+T)=\sum_{n=-\infty}^{\infty} e^{in\omega \tau}\mathcal H_{n}({\bf k}),
\end{align}
where 
\bea 
\mathcal H_{n}({\bf k})=\frac{1}{T}\int_{0}^T e^{-in\omega \tau}\mathcal H({\bf k}, \tau) d\tau=\mathcal H_{-n}^\dg({\bf k}),
\eea
 are the Fourier components and $T=2\pi/\omega$ is the period of the laser light. The corresponding  eigenvectors of the time-periodic Hamiltonian can be written as 
 \bea 
 \ket{\psi_{\alpha}({\bf k}, \tau)}=e^{-i \varepsilon_{\alpha}({\bf k}) \tau}\ket{u_{\alpha}({\bf k}, \tau)},
 \eea
where  $\ket{u_{\alpha}({\bf k}, \tau)}=\ket{u_{\alpha}({\bf k}, \tau+T)}=\sum_{n} e^{in\omega \tau}\ket{u_{\alpha}^n({\bf k})}$ is the time-periodic Floquet-Bloch wave function of magnon and $\varepsilon_{\alpha}({\bf k})$ are the magnon quasienergy modes. We define the Floquet operator  as $\mathcal H_F({\bf k},\tau)=\mathcal H({\bf k},\tau)-i\partial_\tau$.

The corresponding eigenvalue equation is of the form 
\begin{align}
\sum_m \big[\mathcal H_{n-m}({\bf k}) + m\hbar\omega\delta_{n,m}\big]u_{\alpha}^m({\bf k})= \varepsilon_{\alpha}({\bf k})u_{\alpha}^n({\bf k}),
\end{align}
where $n,m$ are integers. The magnon quasienergies $\varepsilon_{\alpha}(\bo)$  can be obtained by diagonalizing the Floquet Hamiltonian
\begin{align}
\mathcal H_{F}= & \begin{pmatrix}
\ddots & \vdots & \vdots & \vdots & \vdots\\
\cdots & \mathcal H_{0}+\hbar\omega & \mathcal H_{1} & \mathcal H_{2} & \cdots\\
\cdots & \mathcal H_{-1} & \mathcal H_{0} & \mathcal H_{1} & \cdots\\
\cdots & \mathcal H_{-2} & \mathcal H_{-1} & \mathcal H_{0}-\hbar\omega & \cdots\\
\vdots & \vdots & \vdots & \vdots & \ddots
\end{pmatrix}.
\label{Floq}
\end{align}
Using the experimentally deduced parameters for  CrI$_3$ \cite{cr}, we obtain the magnon bandwidth  of the undriven system $\mathcal H_{0}$ as $\Delta_{\bf {\Gamma}} = 6(J_1+J_3)S = 19.71$~meV. When the photon energy $\hbar\omega$ is comparable
to $\Delta_{\bf {\Gamma}}$,  different copies of $\mathcal H_{0}$ overlap. However, when the photon energy $\hbar\omega$ is large compared to the magnon bandwidth   $\Delta_{\bf {\Gamma}}$, the Floquet side modes are decoupled. We have considered the latter case.  Thus, the system can be described by the effective time-independent Hamiltonian \cite{pho7} 
\begin{align}
\mathcal H_{eff}(\bo)&=\mathcal H_{0}+\sum_{n\geq 1}\frac{1}{n\hbar\omega}\big[\mathcal H_{n}, \mathcal H_{-n}\big] + \mathcal{O}\Big (\frac{1}{\omega^2}\Big).
\label{effHam}
\end{align}

%
%
%
%
%
%

\end{document}